\documentclass[fp,twocolumn]{jpsj3}
\usepackage{graphicx}
\usepackage{color}
\usepackage{bm}

\title{%
Majorana Zero Mode Induced by a Screw Dislocation on
the Surface of an Iron-based Superconductor
}

\author{%
Naoki Otsuki and Yositake Takane$^{*}$
}

\inst{%
Graduate School of Advanced Science and Engineering,\\
Hiroshima University, Higashihiroshima, Hiroshima 739-8530, Japan
}

\recdate{ \hspace{50mm} }

\abst{%
 We propose a simple scenario to describe a dislocation-induced
Majorana zero mode on the surface of an iron-based superconductor,
using an illustrative model with a cylindrical hole of radius $R$
perpendicular to its top surface.
Topological surface states on the inner surface of the hole form
an effective chiral $p$-wave superconductor.
When the top surface has a perpendicular magnetization,
chiral Majorana modes appear near the circular edge of the hole.
Since the corresponding wavefunctions obey
an antiperiodic boundary condition in the circumferential direction,
a Majorana zero mode at zero energy does not appear.
However, if a screw dislocation is inserted through the hole,
the antiperiodic boundary condition is transformed into a periodic one,
resulting in the appearance of a Majorana zero mode.
This Majorana zero mode remains in the no-hole limit of $R \to 0$.
We confirm this scenario by numerical simulation and effective theory.
A method of creating a Majorana zero mode in the absence of
the surface magnetization is briefly described.
}

\begin{document}
\maketitle

\section{Introduction}

Topological superconductors host Majorana zero modes
at their boundary.~\cite{Kitaev,Alicea,Sato1}
A Majorana zero mode is characterized by the condition that
its creation and annihilation operators are identical
and is allowed to appear at zero energy as a midgap state.
A chiral $p$-wave superconductor is a typical example of
such a topological superconductor.
It has been proposed that a Dirac cone-type two-dimensional electron system
under the proximity effect of $s$-wave superconductivity serves as
an effective chiral $p$-wave topological superconductor.~\cite{fu1}
This proposal opens a way to realizing topological superconductors
in condensed matter systems, including one-dimensional semiconductor nanowires
with spin-orbit interaction~\cite{Lutchyn,Oreg,Mourik}
and magnetic atom chains,~\cite{Nadj-Perge1,Nadj-Perge2,Kim}
under the proximity effect.
A two-dimensional proximity-induced superconductor formed by
the helical surface states of a topological insulator is expected
to host a Majorana zero mode as a vortex bound state.~\cite{fu1,Sau}
Such zero-energy vortex bound states have been observed
on a surface of ${\rm FeSe}_{x}{\rm Te}_{1-x}$
in a magnetic field.~\cite{D-Wang,Machida1,Machida2}

Hu and Zhang~\cite{Hu} have proposed that a Majorana zero mode appears on
a surface of ${\rm FeSe}_{x}{\rm Te}_{1-x}$
in the presence of a screw dislocation.
An interesting feature of this proposal is that the Majorana zero mode can
appear without the application of a magnetic field.
It has been shown that a screw dislocation induces one-dimensional gapless
modes in gapped topological systems~\cite{ran,zhang,teo,imura1,yoshimura,
shiozaki,slager,pauly,Sbierski,hamasaki1,hamasaki2,sakaguchi} as well as
in Weyl and Dirac semimetals.~\cite{imura2,sumiyoshi,chernodub,kodama,huang,
soto-garrido,amitani,zheng}
The Majorana zero mode considered in Ref.~\citen{Hu} should be closely
related to the one-dimensional gapless modes induced by a screw dislocation
in topological insulators.~\cite{ran}

Let us briefly summarize the situation assumed in Ref.~\citen{Hu}.
Previous studies~\cite{Miao,Z-Wang,Xu,Zhang1,Zhang2,Li} revealed that
${\rm FeSe}_{x}{\rm Te}_{1-x}$ is an $s$-wave superconductor
below the transition temperature $T_{\rm c} = 14.5$ K, and above $T_{\rm c}$,
it becomes a strong topological insulator hosting helical surface states,
for which the strong and weak indices~\cite{fu2}
are given as $(\nu_{0},\nu_{x}\nu_{y}\nu_{z}) = (1,001)$.
Below $T_{\rm c}$, the helical surface states form
an effective chiral $p$-wave superconductor under the proximity effect
of $s$-wave superconductivity arising from bulk states.
For the sake of clarity, let us consider a sample of
${\rm FeSe}_{x}{\rm Te}_{1-x}$
with top and bottom surfaces parallel to the $xy$-plane,
into which a screw dislocation with the Burgers vector
$\mib{b}=(0,0,b_{z})$ is inserted.
Here, $\mib{b}=(0,0,b_{z})$ means that the screw dislocation is parallel to
the $z$-axis, and $b_{z}$ unit atomic layers are displaced around it.
Hu and Zhang pointed out that a Majorana zero mode is induced by
the screw dislocation near its end
if the surface where the screw dislocation terminates has
a spontaneous magnetization perpendicular to it.
However, the roles of $(\nu_{x}\nu_{y}\nu_{z}) = (001)$
and $\mib{b}=(0,0,b_{z})$ are not sufficiently clear.
In this paper, we present a simple scenario
that explains the appearance of the Majorana zero mode.
Our scenario clearly shows that the Majorana zero mode appears
owing to the modification of a Bogoliubov quasiparticle (bogolon) wavefunction
induced by the combination of $(\nu_{x}\nu_{y}\nu_{z}) = (001)$
and $\mib{b}=(0,0,b_{z})$ with an odd $b_{z}$.
The condition for the appearance of the Majorana zero mode with respect to
$(\nu_{x}\nu_{y}\nu_{z})$ and $\mib{b}$ is equivalent to
that of the one-dimensional gapless modes.~\cite{comment0}

A comment on the spontaneous surface magnetization in
${\rm FeSe}_{x}{\rm Te}_{1-x}$ is given here.
Its presence is currently unclear.
In fact, the results of some experimental studies~\cite{Zaki,Farhang} suggest
that the spectrum of helical surface states is gapped by surface ferromagnetism
corresponding to the presence of spontaneous surface magnetization
perpendicular to the surface,
while those of other studies~\cite{Z-Wang,Zhang1,Zhang2,Li}
indicate that gapless helical surface states appear on a surface,
suggesting the absence of such surface magnetization.
Nevertheless, we assume that the top surface of a sample of
${\rm FeSe}_{x}{\rm Te}_{1-x}$
has spontaneous surface magnetization perpendicular to the surface.
In the absence of such surface magnetization, we can induce
a constant magnetization on the top surface
by covering it with an insulating ferromagnetic layer.

In the next section, we describe our scenario for the appearance of
the Majorana zero mode, using an illustrative model with a hole
through which a screw dislocation is inserted.
In Sect.~3, we present a tight-binding Hamiltonian on a cubic lattice
for a three-dimensional iron-based superconductor.
We then implement the illustrative model
in the form of a tight-binding Hamiltonian.
In Sect.~4, we numerically determine the spectrum of the resulting model.
The numerical results confirm our scenario.
In Sect.~5, we derive an effective theory in the continuum limit
from the tight-binding Hamiltonian.
In Sects.~6 and 7, we show that the Majorana zero mode appears
under the condition expected in the scenario.
The last section is devoted to a summary and discussion, in which
we briefly describe a method to create a Majorana zero mode
in the absence of spontaneous surface magnetization.

\section{Theoretical Scenario}

To describe our scenario for the appearance of the Majorana zero mode,
we introduce an illustrative model for ${\rm FeSe}_{x}{\rm Te}_{1-x}$
with a fictitious hole of radius $R$ parallel to the $z$-axis
[see Fig.~1(a)].~\cite{imura1,imura2}
Although the hole is not essential for the appearance of
the Majorana zero mode, it helps us to understand this phenomenon.
If $R$ is sufficiently large, helical surface states characterized by
$(\nu_{x}\nu_{y}\nu_{z}) = (001)$ appear on the inner surface of the hole.
Under the proximity effect of $s$-wave superconductivity
arising from bulk states, the helical surface states on the inner surface
form a two-dimensional $s$-wave superconductor,
which can be regarded as an effective chiral $p$-wave superconductor.
Let $\Delta_{0}$ be a pair potential for this superconductor.
On the top surface of the system, the helical surface states are influenced by
a constant magnetization $m_{M}$ perpendicular to the surface;
thus, the surface electron states form
a two-dimensional magnetic insulator if $m_{M}$ is sufficiently large.
Consequently, the effective chiral $p$-wave superconductor on the inner surface
is in contact with the magnetic insulator on the top surface
at a circular edge of the hole of radius $R$.
We expect the appearance of one-dimensional chiral Majorana
modes~\cite{Sato1,fu1} localized near the circular edge
with a circumference of $2\pi R$.
With wave number $k_{\theta}$ in the azimuthal direction,
we assume that their dispersion relation
in the subgap region $-\Delta_{0} < E < \Delta_{0}$ is
\begin{align}
  E(k_{\theta}) = -\hbar v k_{\theta}
\end{align}
with velocity $v$, according to Eq.~(\ref{eq:chiral-disp}).
A wavefunction $\psi(\theta)$ of the chiral Majorana modes obeys
an antiperiodic boundary condition
[i.e., $\psi(\theta) = - \psi(\theta+2\pi)$], as shown in Eq.~(\ref{eq:a-pbc}).
This gives
\begin{align}
  k_{\theta} = \frac{\tilde{n} + \frac{1}{2}}{R}
\end{align}
with an integer $\tilde{n}$.
The energy eigenvalues are given as
\begin{align}
  E = \hbar v \frac{n+ \frac{1}{2}}{R}
\end{align}
with $n = -\tilde{n}-1 = 0, \pm 1, \pm 2, \dots$.
This shows that a Majorana zero mode with $E = 0$ does not appear
since $k_{\theta} = 0$ is not allowed.
The above analysis shows that a Majorana zero mode can appear
if the antiperiodic boundary condition is replaced by a periodic boundary
condition [i.e., $\psi(\theta) = \psi(\theta+2\pi)$],
where $k_{\theta} = 0$ is allowed.
\begin{figure}[btp]
\begin{center}
\includegraphics[height=2.8cm]{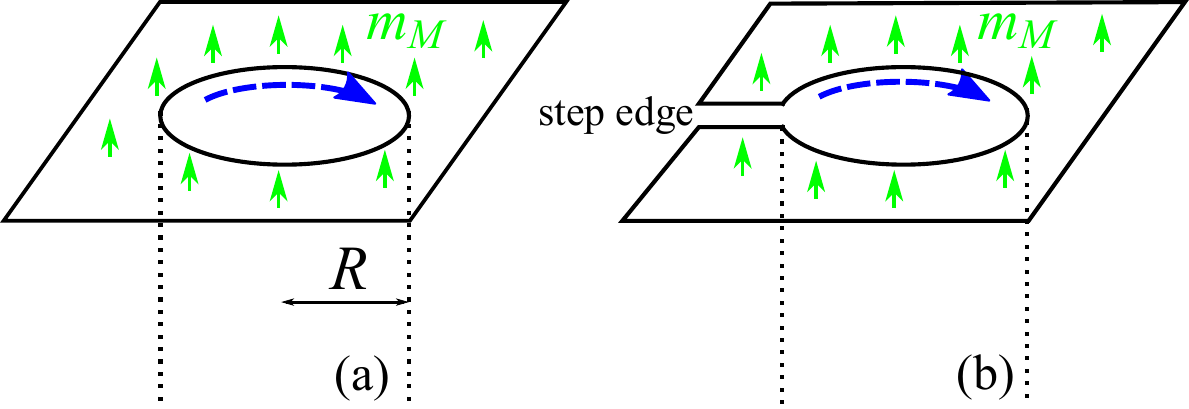}
\end{center}
\caption{
(Color online)
Schematics of the system with a hole of radius $R$.
(a) System with a hole, and (b) that with a hole in the presence of
a screw dislocation that pierces the hole and causes the displacement of
$b_{z}$ unit atomic layers around the hole.
The short arrows represent uniform magnetization on the top surface.
The dashed arrows represent chiral Majorana modes localized near
the circular edge of the hole on the top surface.
}
\end{figure}

Let us introduce a screw dislocation designated by the Burgers vector
$\mib{b}=(0,0,b_{z})$ piercing the hole [see Fig.~1(b)].
The antiperiodic boundary condition is transformed into the periodic one
if $b_{z}$ is an odd integer under the condition that the Dirac point
$(k_{\theta}^{\rm D}, k_{z}^{\rm D})$ of the helical surface states
on the inner surface satisfies $k_{z}^{\rm D} = \frac{\pi}{a}$,
which is guaranteed by the weak indices $(001)$ in our setup.
The transformation is essentially based on the weak indices $(001)$ and
$\mib{b}=(0,0,b_{z})$, as explained in Sect~5.
Under the periodic boundary condition,
the eigenvalues of the chiral Majorana modes are
\begin{align}
      \label{eq:spect-with-MZM}
  E = \hbar v \frac{n}{R}
\end{align}
with $n = 0, \pm 1, \pm 2, \dots$,
showing the appearance of a Majorana zero mode at zero energy.
As $R$ is decreased to $0$, each positive (negative) eigenvalue increases
(decreases) and eventually reaches the upper (lower) end of the subgap region
$-\Delta_{0} < E < \Delta_{0}$,
where the corresponding chiral Majorana mode becomes a bulk state.
Thus, the chiral Majorana modes with a nonzero eigenvalue disappear
in the no-hole limit of $R \to 0$.
In contrast, the Majorana zero mode at $E = 0$ remains in the same limit.
We thus conclude that the Majorana zero mode appears
when a screw dislocation with an odd $b_{z}$ pierces the system,
regardless of the presence or absence of the hole.

This scenario is verified in Sect.~4 by numerical simulation
and in Sects.~6 and 7 by means of effective theory in the continuum limit.

\section{Tight-binding Model}

Using a tight-binding model for $\mathbb{Z}_{2}$ topological
insulators,~\cite{liu} we present a tight-binding model for
${\rm FeSe}_{x}{\rm Te}_{1-x}$ on the cubic lattice
in a rectangular parallelepiped shape of
volume $N_{x} \times N_{y} \times N_{z}$ with the lattice constant $a$.
The indices $l$, $m$, and $n$ are used to specify lattice sites
in the $x$-, $y$-, and $z$-directions, respectively, where
$1 \le l \le N_{x}$, $1 \le m \le N_{y}$, and $1 \le n \le N_{z}$.
We apply a periodic boundary condition in the $x$- and $y$-directions.
To describe a $\mathbb{Z}_{2}$ topological insulator, we must consider
orbital and spin degrees of freedom described by
$\tau = 1, 2$ and $\sigma = \uparrow, \downarrow$, respectively.
Since the bogolon states in a superconductor are described in the Nambu space
consisting of electron and hole spaces, we must also consider
electron--hole degrees of freedom described by $\eta = e, h$.
The four-component state vectors for the $(l,m,n)$th site
defined in the electron and hole spaces are expressed as
\begin{align}
  |l,m,n \rangle^{e}
 &  = \Bigl[ |l,m,n \rangle_{1\uparrow}^{e} \hspace{0.8mm}
             |l,m,n \rangle_{2\uparrow}^{e} \hspace{0.8mm}
             |l,m,n \rangle_{1\downarrow}^{e} \hspace{0.8mm}
             |l,m,n \rangle_{2\downarrow}^{e}
      \Bigr] ,
      \\
  |l,m,n \rangle^{h}
 &  = \Bigl[ |l,m,n \rangle_{1\uparrow}^{h} \hspace{0.8mm}
             |l,m,n \rangle_{2\uparrow}^{h} \hspace{0.8mm}
             |l,m,n \rangle_{1\downarrow}^{h} \hspace{0.8mm}
             |l,m,n \rangle_{2\downarrow}^{h}
      \Bigr] ,
\end{align}
respectively.
With the basis consisting of $|l,m,n \rangle^{\eta}$ and
${}^{\eta}\langle l,m,n| \equiv (|l,m,n \rangle^{\eta})^{\dagger}$,
the Bogoliubov--de Gennes Hamiltonian for this system is given by
\begin{align}
        \label{eq:H_BdG}
  H_{\rm BdG}
  = \left[
      \begin{array}{cc}
         H_{e} & \Delta \\
         \Delta^{\dagger} & H_{h}
      \end{array}
    \right] ,
\end{align}
where $H_{e}$ and $H_{h}$ describe electrons and holes, respectively,
in a $\mathbb{Z}_{2}$ topological insulator and
\begin{align}
    \label{eq:Delta}
   \Delta
 & = \sum_{l=1}^{N_{x}}\sum_{m=1}^{N_{y}}\sum_{n=1}^{N_{z}}
       |l,m,n \rangle^{e}
       \left[ \begin{array}{cc}
                0_{2 \times 2} & \Delta_{0}\tau_{0} \\
                -\Delta_{0}\tau_{0} & 0_{2 \times 2}
              \end{array}
       \right]
       {}^{h}\langle l,m,n|
\end{align}
represents the pair potential
with $\tau_{0}$ being the $2 \times 2$ identity matrix.
Here, $H_{\eta}$ with $\eta = e, h$ is decomposed as
$H_{\eta} = H_{\eta}^{\rm d}+H_{\eta}^{x}+H_{\eta}^{y}+H_{\eta}^{z}$
with
\begin{align}
    \label{eq:H_d}
   H_{\eta}^{\rm d}
 & = \sum_{l=1}^{N_{x}}\sum_{m=1}^{N_{y}}\sum_{n=1}^{N_{z}}
     |l,m,n \rangle^{\eta} h_{\eta}^{d} {}^{\eta}\langle l,m,n| ,
         \\
    \label{eq:H_x}
   H_{\eta}^{x}
 & = \sum_{l=1}^{N_{x}}\sum_{m=1}^{N_{y}}\sum_{n=1}^{N_{z}}
     |l+1,m,n \rangle^{\eta} h_{\eta}^{x} {}^{\eta}\langle l,m,n|
     + {\rm h.c.} ,
        \\
    \label{eq:H_y}
   H_{\eta}^{y}
 & = \sum_{l=1}^{N_{x}}\sum_{m=1}^{N_{y}}\sum_{n=1}^{N_{z}}
     |l,m+1,n \rangle^{\eta} h_{\eta}^{y} {}^{\eta}\langle l,m,n|
     + {\rm h.c.} ,
        \\
    \label{eq:H_z}
   H_{\eta}^{z}
 & = \sum_{l=1}^{N_{x}}\sum_{m=1}^{N_{y}}\sum_{n=1}^{N_{z}-1}
     |l,m,n+1 \rangle^{\eta} h_{\eta}^{z} {}^{\eta}\langle l,m,n|
     + {\rm h.c.} ,
\end{align}
where
\begin{align}
   h_{e}^{d} & = - h_{h}^{d}
   = \left[ 
       \begin{array}{cc}
         \tilde{m}\tau_{z}-\mu\tau_{0} & 0_{2 \times 2} \\
         0_{2 \times 2} & \tilde{m}\tau_{z}-\mu\tau_{0}
       \end{array}
     \right] ,
       \\
   h_{e}^{x} & = - (h_{h}^{x})^{*}
   = \left[ 
       \begin{array}{cc}
         -m_{\parallel}\tau_{z} & \frac{iA}{2}\tau_{x} \\
         \frac{iA}{2}\tau_{x} & -m_{\parallel}\tau_{z}
       \end{array}
     \right] ,
       \\
   h_{e}^{y} & = - (h_{h}^{y})^{*}
   = \left[ 
       \begin{array}{cc}
         -m_{\parallel}\tau_{z} & \frac{A}{2}\tau_{x} \\
         -\frac{A}{2}\tau_{x} & -m_{\parallel}\tau_{z}
       \end{array}
     \right] ,
       \\
   h_{e}^{z} & = - (h_{h}^{z})^{*}
   = \left[ 
       \begin{array}{cc}
         -m_{\perp}\tau_{z} + \frac{iA}{2}\tau_{x} & 0_{2 \times 2} \\
         0_{2 \times 2} & -m_{\perp}\tau_{z} - \frac{iA}{2}\tau_{x}
       \end{array}
     \right] .
\end{align}
Here, $\tilde{m} = m_{0} + 4m_{\parallel} + 2 m_{\perp}$
and $\mu$ is the chemical potential.
The Pauli matrices $\tau_{q}$ ($q \in x, y, z$) are defined
together with $\tau_{0}$ in the orbital space.
The strong and weak indices $(\nu_{0}, \nu_{x}\nu_{y}\nu_{z})$
are determined by the parameters $m_{0}$, $m_{\parallel}$, and $m_{\perp}$.
We set $m_{\perp}/m_{\parallel} = -1$ and $m_{0}/m_{\parallel} = 2$
throughout this paper, which ensures that both $H_{e}$ and $H_{h}$
describe a strong topological insulator with
$(\nu_{0}, \nu_{x}\nu_{y}\nu_{z}) = (1,001)$.~\cite{imura3}
The bulk spectrum of $H_{e}$ is $E(\mib{k}) = \pm \sqrt{\Xi({\mib k})}$, where
${\mib k} = (k_{x}, k_{y}, k_{z})$ and
\begin{align}
 \Xi({\mib k})
     & = [\tilde{m} - 2m_{\parallel}\left(\cos(k_{x}a) + \cos(k_{y}a)\right)
            - 2m_{\perp}\cos(k_{z}a)]^{2}
     \nonumber \\
     & \hspace{-2mm}
         + A^{2}
           \left(\sin^{2}(k_{x}a)+\sin^{2}(k_{y}a)+\sin^{2}(k_{z}a)\right) .
\end{align}
In our case,
with $m_{0}/m_{\parallel} = 2$ and $m_{\perp}/m_{\parallel} = - 1$,
the energy gap defined as $E_{\rm g} = \min_{\mib k}\{\sqrt{\Xi({\mib k})}\}$
is $E_{\rm g}/m_{\parallel} = 2$.

Our model under the periodic boundary condition in the $x$- and $y$-directions
can describe an effective chiral $p$-wave superconductor
appearing on the top and bottom surfaces.
In ${\rm FeSe}_{x}{\rm Te}_{1-x}$,
a topologically nontrivial bulk band coexists with trivial bulk bands
that induce $s$-wave superconductivity.~\cite{Zhang1,Zhang2}
Since our attention is focused on the effective chiral $p$-wave superconductor
induced by the nontrivial bulk band,
the trivial bulk bands are neglected in the model
and only their proximity effect is included
in terms of the pair potential $\Delta$ describing $s$-wave pairing.
Although $s_{\pm}$-wave pairing has been proposed
in several studies,~\cite{Mazin,Kuroki,Hanaguri}
we use the $s$-wave pairing model
because the details of the pairing are not important for our analysis.

The uniform magnetization $\mib{M}=(0,0,m_{M})$ on the top surface
is included in $H_{\rm BdG}$
by adding $V_{\eta}^{M}$ to $h_{\eta}^{d}$ with
\begin{align}
 V_{e}^{M} & = - V_{h}^{M}
   = \delta_{n, N_{z}}
     \left[ 
       \begin{array}{cc}
         m_{M}\tau_{0} & 0_{2 \times 2} \\
         0_{2 \times 2} & -m_{M}\tau_{0}
       \end{array}
     \right] .
\end{align}
\begin{figure}[btp]
\begin{center}
\includegraphics[height=5.5cm]{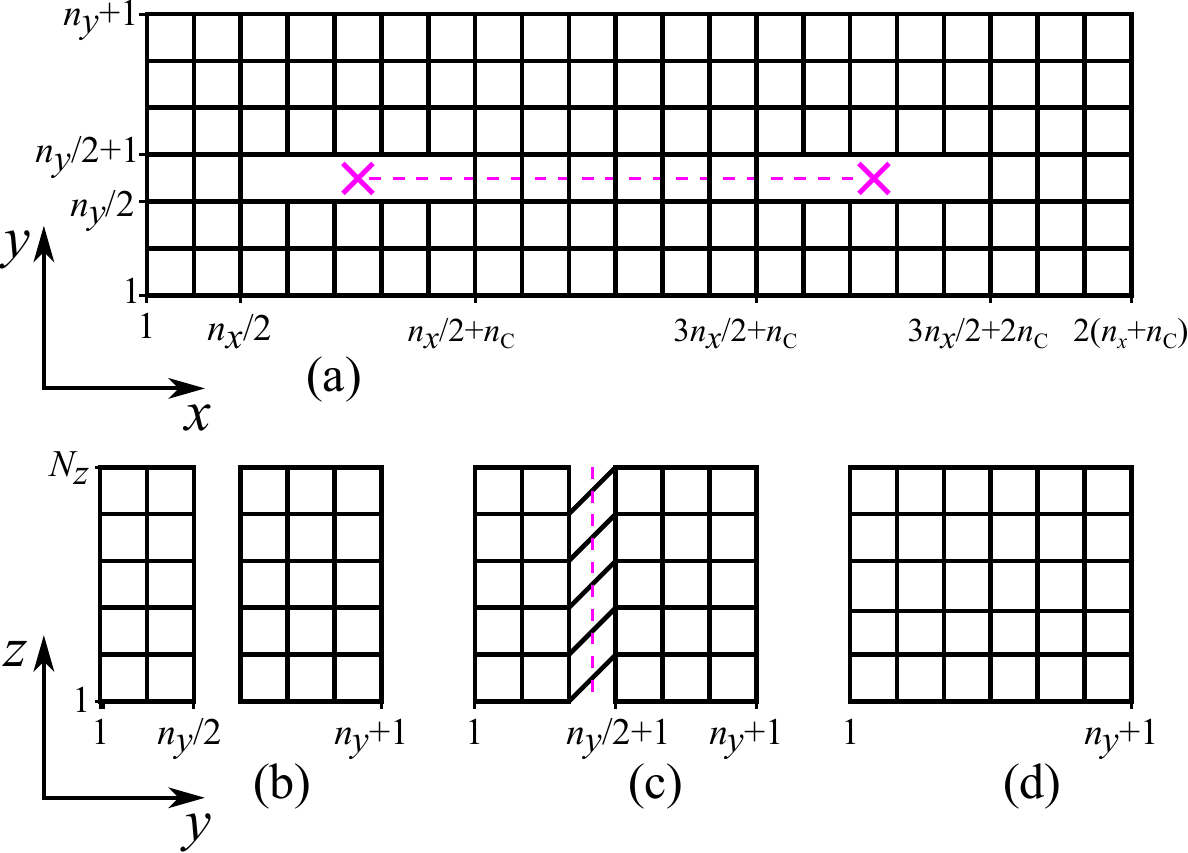}
\end{center}
\caption{
(Color online)
Lattice system with two cuts and a pair of screw dislocations,
where the crosses represent the dislocation centers,
around each of which one unit atomic layer (i.e., $b_{z} = 1$) is displaced.
Each solid line between two neighboring sites represents
a hopping term in $H_{x}$, $H_{y}$, or $H_{z}$
directly connecting the two sites.
The dotted lines represent
the slip plane across which the hopping terms in $H$ are modified.
(a) Top view of the system.
(b) Side view of the cut regions of
$\frac{n_{x}}{2}+1 \le l \le \frac{n_{x}}{2}+n_{\rm C}-1$
and $\frac{3n_{x}}{2}+n_{\rm C}+1 \le l \le \frac{3n_{x}}{2}+2n_{\rm C}-1$.
(c) Side view of the modified region of
$\frac{n_{x}}{2}+n_{\rm C} \le l \le \frac{3n_{x}}{2}+n_{\rm C}$.
(d) Side view of the unmodified regions of $1 \le l \le \frac{n_{x}}{2}$
and $\frac{3n_{x}}{2}+2n_{\rm C} \le l \le N_{x}$.
}
\end{figure}
Let us make a pair of cuts parallel to the $xz$-plane
in our tight-binding model and introduce a screw dislocation
parallel to the $z$-axis at the center of each cut [see Figs.~2(a)--2(d)].
Each cut of width $n_{\rm C}$
corresponds to the hole in the illustrative model.
That is, our tight-binding model has two holes.
We assume that the number of sites in the $x$-direction is
$N_{x} = 2\left(n_{x}+n_{\rm C}\right)$ and that in the $y$-direction
is $N_{y} = n_{y}+1$, where $n_{x}$ and $n_{y}$ are even integers
and $n_{\rm C}$ is an odd integer.
The two cuts are made by removing the hopping terms in $H_{\eta}^{y}$
between $m = \frac{n_{y}}{2}$ and $m = \frac{n_{y}}{2}+1$ in the regions
$\frac{n_{x}}{2}+1 \le l \le \frac{n_{x}}{2}+n_{\rm C}-1$ and
$\frac{3n_{x}}{2}+n_{\rm C}+1 \le l \le \frac{3n_{x}}{2}+2n_{\rm C}-1$
[see Fig.~2(b)].
The Burgers vectors of the two screw dislocations are
$\mib{b}_{1} = (0,0,b_{z})$ and $\mib{b}_{2} = (0,0,-b_{z})$.
Figure~2 shows the case of $n_{x} = n_{y} = 6$, $n_{\rm C} = 5$,
and $N_{z} = 6$ with $b_{z} = 1$.
The case of $n_{\rm C} = 1$ and $b_{z} \neq 0$ corresponds to the system
with the two screw dislocations without cuts.
The reason for introducing the two dislocations is to apply
the periodic boundary condition in the $x$- and $y$-directions.
The two screw dislocations modify the hopping terms in $H_{\eta}^{y}$
between $m = \frac{n_{y}}{2}$ and $m = \frac{n_{y}}{2}+1$ in the region
$\frac{n_{x}}{2}+n_{\rm C} \le l \le \frac{3n_{x}}{2}+n_{\rm C}$
so that each term connects two different layers in the $z$-direction
[see Fig.~2(c)].~\cite{imura1,imura2}
In the presence of the two cuts and two dislocations, $H_{\eta}^{y}$ becomes
\begin{align}
      H_{\eta}^{y}
 &  = \sum_{l=1}^{N_{x}}
      \sum_{\substack{m=1 \\ \left(m \neq \frac{n_{y}}{2}\right)}}^{N_{y}}
      \sum_{n=1}^{N_{z}}
      |l,m+1,n \rangle^{\eta} h_{\eta}^{y} {}^{\eta}\langle l,m,n| + {\rm h.c.}
      \nonumber \\
 & \hspace{-2mm}
    + \sum_{l=1}^{\frac{n_{x}}{2}}\sum_{n=1}^{N_{z}}
      |l,\frac{n_{y}}{2}+1,n \rangle^{\eta} h_{\eta}^{y}
                             {}^{\eta}\langle l,\frac{n_{y}}{2},n|
      + {\rm h.c.}
      \nonumber \\
 & \hspace{-2mm}
    + \sum_{l=\frac{n_{x}}{2}+n_{\rm C}}^{\frac{3n_{x}}{2}+n_{\rm C}}
      \sum_{n=1}^{N_{z}-b_{z}}
      |l,\frac{n_{y}}{2}+1,n+b_{z} \rangle^{\eta} h_{\eta}^{y}
                             {}^{\eta}\langle l,\frac{n_{y}}{2},n|
      + {\rm h.c.}
      \nonumber \\
 & \hspace{-2mm}
    + \sum_{l=\frac{3n_{x}}{2}+2n_{\rm C}}^{N_{x}}\sum_{n=1}^{N_{z}}
      |l,\frac{n_{y}}{2}+1,n \rangle^{\eta} h_{\eta}^{y}
                             {}^{\eta}\langle l,\frac{n_{y}}{2},n|
      + {\rm h.c.} 
\end{align}
A step edge of height $b_{z}$ is induced by the screw dislocations
on the top and bottom surfaces [see Fig.~2(c)].

\section{Numerical Results}

Using the two-hole lattice system described in Sect.~3,
we numerically compute the energy eigenvalues in this system
in the presence or absence of screw dislocations with $b_{z} = 1$.
The system size is fixed as
$N_{x} \times N_{y} \times N_{z} = 74 \times 15 \times 12$ with $n_{y} = 14$.
We study the four cases of $n_{\rm C} = 23$ and $n_{x} = 14$,
$n_{\rm C} = 15$ and $n_{x} = 22$, $n_{\rm C} = 11$ and $n_{x} = 26$,
and $n_{\rm C} = 1$ and $n_{x} = 36$, in each of which $N_{x} = 74$.
We set $m_{0}/m_{\parallel} = 2$ and $m_{\perp}/m_{\parallel} = - 1$,
so that $(\nu_{x}\nu_{y}\nu_{z}) = (001)$.
The other parameters are $A/m_{\parallel} = 1$,
$\Delta_{0}/m_{\parallel} = 0.6$, and $m_{M}/m_{\parallel} = 1.2$.
The chemical potential is fixed at $\mu/m_{\parallel} = 0$
unless otherwise noted.
Our attention is focused on chiral Majorana modes, which should appear near
the edge of each hole on the top surface.
Half of the superconducting energy gap in this system is equal to $\Delta_{0}$;
thus, the chiral Majorana modes appear
in the subgap region of $-0.6 < E/m_{\parallel} < 0.6$.
In the following, we focus on the eigenvalues in this region.
\begin{figure}[tbp]
\begin{tabular}{cc}
\begin{minipage}{0.5\hsize}
\begin{center}
\hspace{-10mm}
\includegraphics[height=5.3cm]{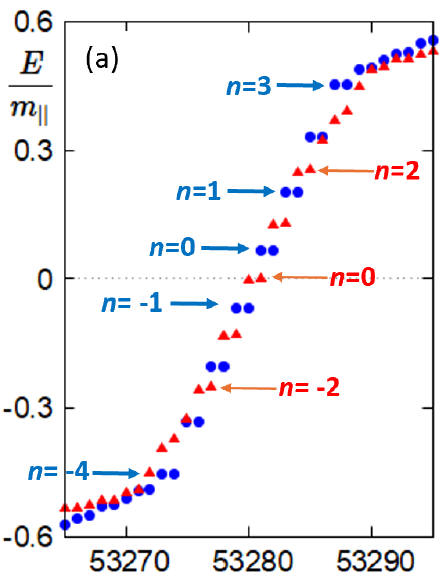}
\end{center}
\end{minipage}
\begin{minipage}{0.5\hsize}
\begin{center}
\hspace{-5mm}
\includegraphics[height=5.3cm]{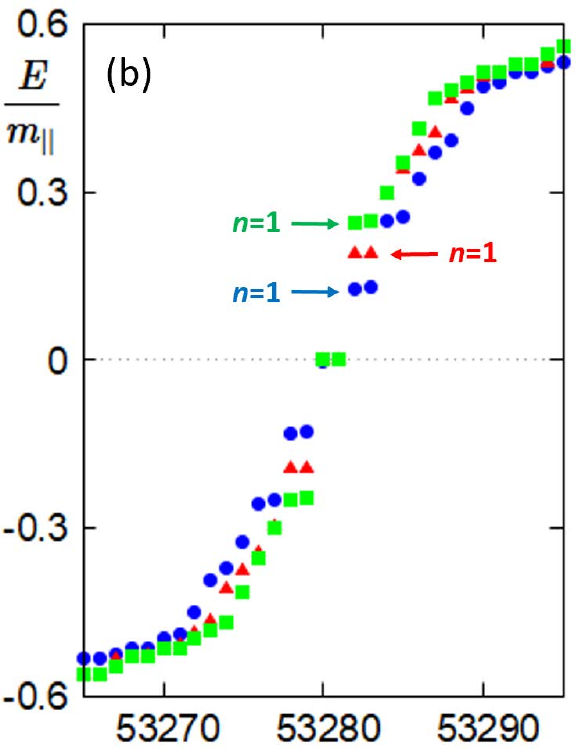}
\end{center}
\end{minipage}
\end{tabular}
\begin{tabular}{cc}
\begin{minipage}{0.5\hsize}
\begin{center}
\hspace{-10mm}
\includegraphics[height=5.3cm]{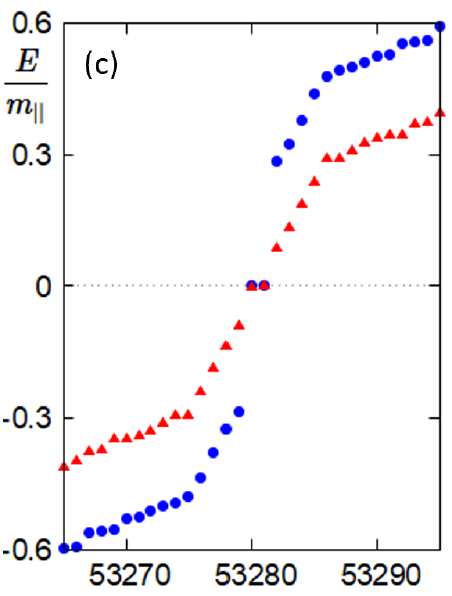}
\end{center}
\end{minipage}
\begin{minipage}{0.5\hsize}
\begin{center}
\hspace{-5mm}
\includegraphics[height=5.3cm]{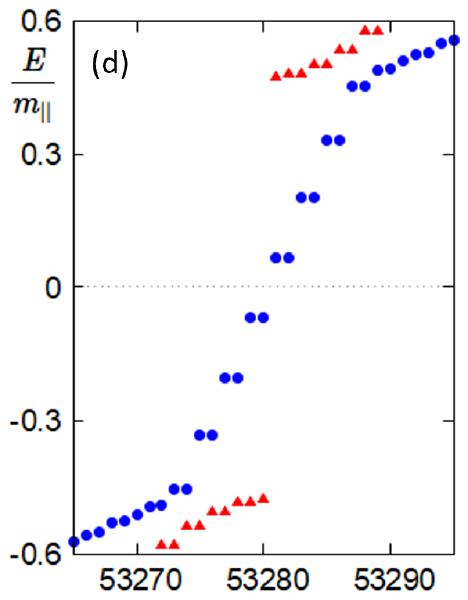}
\end{center}
\end{minipage}
\end{tabular}
\caption{
(Color online)
The energy eigenvalues in the subgap region are shown
in increasing order from left to right.
(a) Eigenvalues in the case with the cuts of width $n_{\rm C} = 23$
in the absence (circles) and presence (triangles) of
the screw dislocations.
(b) Eigenvalues in the cases with the cuts of width
$n_{\rm C} = 23$ (circles), $15$ (triangles), and $11$ (squares)
in the presence of the screw dislocations.
(c) Eigenvalues in the case without the cuts in the presence
of the screw dislocations
at $\mu/m_{\parallel} = 0$ (circles) and  $0.5$ (triangles).
(d) Eigenvalues in the case without the cuts (triangles)
and in the case with the cuts of width $n_{\rm C} = 23$ (circles)
in the absence of the screw dislocations.
}
\end{figure}

Figure~3(a) shows the energy eigenvalues in the system with two cuts
of width $n_{\rm C} = 23$
in the absence and presence of the screw dislocations.
In the absence of the screw dislocations, we observe
the eight doubly degenerate eigenvalues labeled by
$n \in \{0, \pm 1, \pm 2, \pm3, -4\}$ in the subgap region.
They must correspond to the chiral Majorana modes
because the distances between neighboring eigenvalues are nearly equal
in accordance with the spectrum $E = \hbar v (n+\frac{1}{2})/R$
expected in our theoretical scenario described in Sect.~2.~\cite{comment1}
In the presence of the screw dislocations, we observe
the five doubly degenerate eigenvalues labeled by $n \in \{0, \pm 1, \pm 2\}$
in the subgap region.
Again, the distances between neighboring eigenvalues are nearly equal
in accordance with the expected spectrum $E = \hbar v n/R$,
showing that they correspond to the chiral Majorana modes.
We observe that the doubly degenerate eigenvalue at $E = 0$, corresponding to
the Majorana zero modes,
appears only in the presence of the screw dislocations.
Figure~3(b) shows the energy eigenvalues in the presence of
the screw dislocations in the cases of $n_{\rm C} = 23, 15$, and $11$.
The doubly degenerate eigenvalue labeled by $n = 1$ increases
with decreasing $n_{\rm C}$.
Figure~3(c) shows the energy eigenvalues in the presence of
the screw dislocations in the no-hole limit of $n_{\rm C} = 1$
at $\mu/m_{\parallel} = 0$ and  $0.5$.
We find only one doubly degenerate eigenvalue at $E = 0$,
showing that the chiral Majorana modes disappear
except for the Majorana zero modes at $E = 0$.
The above results are consistent with our scenario described in Sect~2,
indicating that the Majorana zero modes in the no-hole limit share
the origin with those in the systems with the holes.

Here, we comment on several subgap states that cannot be assigned to
the chiral Majorana modes.
They are located near the edges $\pm \Delta_{0}$ of the subgap region.
Figure~3(d) shows the energy eigenvalues in the cases without the cuts
($n_{\rm C} = 1$) and with the cuts ($n_{\rm C} = 23$)
in the absence of the screw dislocations.
These subgap states are localized near the top surface (data not shown);
thus, they should be considered as trivial subgap states
induced by surface magnetization.
From Fig.~3(d), we observe  that the number of the trivial subgap states
is increased by the cuts.
This is ascribed to the hybridization of some chiral Majorana modes
with trivial subgap states.
We also observe from Figs.~3(a) and 3(d) that more trivial subgap states appear
in the presence of the screw dislocations than in their absence
in the case of the cuts ($n_{\rm C} = 23$).
A reasonable explanation for this is that some trivial subgap states are
localized near the step edge [see Fig.~2(c)]
and are hybridized with some chiral Majorana modes.
This explains the observation that the eigenvalues labeled by $n = \pm 3$
cannot be unambiguously identified in Fig.~3(a)
in the presence of the screw dislocations.

\section{Effective Hamiltonian}

Let us consider a system described by $H_{\rm BdG}$, which is infinitely long
in the $z$-direction as well as in the other two directions,
and has a hole of radius $R$ along the $z$-axis.
The bogolon states on the cylindrical inner surface of circumference $2\pi R$
form the effective chiral $p$-wave superconductor.
To describe this, we derive an effective Hamiltonian for helical surface
states of electrons starting from $H_{e}$.~\cite{imura4,takane}
The Fourier transform of $H_{e}$ is
\begin{align}
 \hspace{-8mm}
 & H_{e}(\mib{k}) =
        \nonumber \\
 & \left[ \begin{array}{cc}
            m(\mib{k})\tau_{z}+A\sin(k_{z}a)\tau_{x}
              & A p_{-}(k_{x},k_{y}) \tau_{x} \\
            A p_{+}(k_{x},k_{y})\tau_{x}
              & m(\mib{k})\tau_{z}-A\sin(k_{z}a)\tau_{x}
          \end{array}
   \right]
\end{align}
with
\begin{align}
  m(\mib{k})
  & = \tilde{m} - 2m_{\parallel}\left(\cos(k_{x}a) + \cos(k_{y}a)\right)
          \nonumber \\
  &  \hspace{8mm} - 2m_{\perp}\cos(k_{z}a) ,
        \\
  p_{\pm}(k_{x},k_{y})
  & = \sin(k_{x}a) \pm i\sin(k_{y}a) .
\end{align}
If the system has a plane surface parallel to the $xz$- or $yz$-plane,
the Dirac point of the surface states is located at
$(k_{x/y}^{\rm D}, k_{z}^{\rm D}) = (0,\frac{\pi}{a})$
because $(\nu_{x}\nu_{y}\nu_{z}) = (001)$.

We use $z$ and an azimuthal angle $\theta$ together with $R$
to specify a point on the cylindrical inner surface.
As described in Appendix~A,
the effective Hamiltonian $H_{e}^{\rm eff}$ for the helical surface states
of electrons is obtained as
\begin{align}
    \label{eq:H_e_eff}
 H_{e}^{\rm eff}
 = \left[ \begin{array}{cc}
            0 & \mathcal{A}_{\theta}\hat{k}_{\theta}
                -i\mathcal{A}\left(\hat{k}_{z}-\frac{\pi}{a}\right) \\
            \mathcal{A}_{\theta}\hat{k}_{\theta}
            +i\mathcal{A}\left(\hat{k}_{z}-\frac{\pi}{a}\right) & 0
          \end{array}
   \right] ,
\end{align}
where $\mathcal{A}_{\theta}=Aa+\frac{m_{\parallel}a^{2}}{R}$,
$\mathcal{A}=Aa$, and
\begin{align}
    \label{eq:def-k_theta-z}
  \hat{k}_{\theta} = \frac{1}{iR}\frac{\partial}{\partial \theta} ,
  \hspace{10mm}
  \hat{k}_{z} = \frac{1}{i}\frac{\partial}{\partial z} .
\end{align}
The basis vectors used to express $H_{e}^{\rm eff}$ are
\begin{align}
     \label{eq:def-basis-v1}
  |\varphi_{e 1}(\theta)\rangle
  & = \frac{1}{2}
      \left[ \begin{array}{c}
               -\left[ \begin{array}{c}
                         1 \\
                         i
                       \end{array}
                \right]
                e^{-i\frac{\theta}{2}} \\
                \left[ \begin{array}{c}
                         1 \\
                         i
                       \end{array}
                \right]
                e^{i\frac{\theta}{2}}
             \end{array}
      \right] ,
         \\
     \label{eq:def-basis-v2}
  |\varphi_{e 2}(\theta)\rangle
  & = \frac{1}{2}
      \left[ \begin{array}{c}
                \left[ \begin{array}{c}
                         1 \\
                         -i
                       \end{array}
                \right]
                e^{-i\frac{\theta}{2}} \\
                \left[ \begin{array}{c}
                         1 \\
                         -i
                       \end{array}
                \right]
                e^{i\frac{\theta}{2}}
             \end{array}
      \right] .
\end{align}
The effective Hamiltonian $H_{h}^{\rm eff}$ for the helical surface states
of holes is also obtained as
\begin{align}
    \label{eq:H_h_eff}
 H_{h}^{\rm eff}
 = \left[ \begin{array}{cc}
            0 & \mathcal{A}_{\theta}\hat{k}_{\theta}
                +i\mathcal{A}\left(\hat{k}_{z}-\frac{\pi}{a}\right) \\
            \mathcal{A}_{\theta}\hat{k}_{\theta}
            -i\mathcal{A}\left(\hat{k}_{z}-\frac{\pi}{a}\right) & 0
          \end{array}
   \right] .
\end{align}
The basis vectors used to express $H_{h}^{\rm eff}$ are
\begin{align}
     \label{eq:def-basis-v3}
  |\varphi_{h 1}(\theta)\rangle
  & = \frac{1}{2}
      \left[ \begin{array}{c}
               -\left[ \begin{array}{c}
                         1 \\
                         -i
                       \end{array}
                \right]
                e^{i\frac{\theta}{2}} \\
                \left[ \begin{array}{c}
                         1 \\
                         -i
                       \end{array}
                \right]
                e^{-i\frac{\theta}{2}}
             \end{array}
      \right] ,
         \\
     \label{eq:def-basis-v4}
  |\varphi_{h 2}(\theta)\rangle
  & = \frac{1}{2}
      \left[ \begin{array}{c}
                \left[ \begin{array}{c}
                         1 \\
                         i
                       \end{array}
                \right]
                e^{i\frac{\theta}{2}} \\
                \left[ \begin{array}{c}
                         1 \\
                         i
                       \end{array}
                \right]
                e^{-i\frac{\theta}{2}}
             \end{array}
      \right] .
\end{align}

As a result, the effective Bogoliubov--de Gennes Hamiltonian is written as
\begin{align}
 H_{\rm BdG}^{\rm eff}
 = \left[ \begin{array}{cc}
            H_{e}^{\rm eff} - \mu & \Delta^{\rm eff} \\
            {\Delta^{\rm eff}}^{\dagger} & H_{h}^{\rm eff} + \mu
          \end{array}
   \right]
\end{align}
with
\begin{align}
    \label{eq:Delta_eff}
 \Delta^{\rm eff}
 = \left[ \begin{array}{cc}
            0 & -\Delta_{0} \\
            \Delta_{0} & 0
          \end{array}
   \right] .
\end{align}
A derivation of $\Delta^{\rm eff}$ is also described in Appendix~A.

We consider what boundary condition should be imposed on an eigenfunction
of $H_{\rm BdG}^{\rm eff}$ with respect to $\theta$.~\cite{takane}
Let us introduce an eigenfunction of $H_{\eta}^{\rm eff}$ ($\eta = e$ or $h$)
satisfying $H_{\eta}^{\rm eff}\psi_{\eta}(z,\theta) = E \psi_{\eta}(z,\theta)$
with
\begin{align}
  \psi_{\eta}(z,\theta) = \left[\begin{array}{c}
                             a_{\eta}(z,\theta) \\ b_{\eta}(z,\theta)
                             \end{array}
                       \right] .
\end{align}
The corresponding state in the original four-component electron or hole space
is expressed as
\begin{align}
    \label{eq:rep-4comp}
  |\Psi_{\eta}(z,\theta)\rangle
  =  a_{\eta}(z,\theta)|\varphi_{\eta 1}(\theta)\rangle
   + b_{\eta}(z,\theta)|\varphi_{\eta 2}(\theta)\rangle .
\end{align}
This state must satisfy
\begin{align}
    \label{eq:rep-4comp-pbc}
  |\Psi_{\eta}(z,\theta)\rangle = |\Psi_{\eta}(z,\theta+2\pi)\rangle .
\end{align}
Since the basis vectors obey
\begin{align}
  |\varphi_{\eta i}(\theta)\rangle = - |\varphi_{\eta i}(\theta+2\pi)\rangle
\end{align}
for $i = 1$ and $2$, as is clear from Eqs.~(\ref{eq:def-basis-v1}),
(\ref{eq:def-basis-v2}), (\ref{eq:def-basis-v3}), and (\ref{eq:def-basis-v4}),
we understand that Eq.~(\ref{eq:rep-4comp-pbc}) requires
the antiperiodic boundary condition
\begin{align}
     \label{eq:a-pbc}
  \psi_{\eta}(z,\theta) = - \psi_{\eta}(z,\theta+2\pi) .
\end{align}
From this analysis,
we conclude that an eigenfunction of $H_{\rm BdG}^{\rm eff}$ must satisfy
the antiperiodic boundary condition with respect to $\theta$.
This can be understood as a result of the spin-momentum locking of 
the helical surface states~\cite{zhang,takane,rosenberg}.

\section{Chiral Majorana Modes}

In the illustrative model,
which occupies the semi-infinite region of $z \le 0$,
the cylindrical hole is connected to the top surface at $z = 0$,
where the magnetization $m_{M}$ is perpendicular to the surface.
Our purpose in this section is to show that the chiral Majorana modes appear
near the circular edge of the hole on the top surface.
The bogolon states on the cylindrical inner surface
penetrate onto the top surface.
To describe this in a simple way, we use $H_{\rm BdG}^{\rm eff}$
derived for the infinitely long hole system
and consider the inner surface of the hole in the region $z > 0$
as the top surface in the original setup.
To simulate the magnetization on the top surface,
a constant magnetization perpendicular to the inner surface is added
to $H_{\rm BdG}^{\rm eff}$ in the region of $z > 0$.
This is done by adding $U_{M}$ to $H_{\rm BdG}^{\rm eff}$, where
\begin{align}
 U_{M}
 = \left[ \begin{array}{cc}
            m_{M}^{\rm eff}(z)\tau_{z} & 0_{2 \times 2} \\
            0_{2 \times 2} & -m_{M}^{\rm eff}(z)\tau_{z}
          \end{array}
   \right]
\end{align}
with
\begin{align}
  m_{M}^{\rm eff}(z)
  = \left\{ \begin{array}{cc}
              m_{M}^{\rm eff} & {\rm for} \hspace{3mm} z > 0 \\
              0 & {\rm for} \hspace{3mm} z \le 0
            \end{array}
    \right. .
\end{align}
The effective magnetization $m_{M}^{\rm eff}$ is smaller than $m_{M}$
depending on the penetration depth of the surface states
[see Eq.~(\ref{eq:eigenf-penet})]

We expect that this model effectively describes the low-energy properties of
the illustrative model.
Our attention is focused on the case of $m_{M}^{\rm eff} > \Delta_{0}$,
where the energy gap of the bogolon states in the region $z > 0$ is
determined by $m_{M}^{\rm eff}$.
In other words, the bogolon states in the region of $z > 0$
form a magnetic insulator.

Solving the eigenvalue equation for $H_{\rm BdG}^{\rm eff}+U_{M}$,
we show that one-dimensional chiral Majorana modes localized near $z = 0$
appear when
\begin{align}
     \label{eq:condition-mu}
  \sqrt{{m_{M}^{\rm eff}}^{2}-\Delta_{0}^{2}} > |\mu| ,
\end{align}
where $m_{M}^{\rm eff} > \Delta_{0}$ is implicitly assumed.
It is convenient to introduce a trial function,
\begin{align}
      \label{eq:trial-eigenf}
   \phi(z,\theta)
   = e^{i\left(k_{z}+\frac{\pi}{a}\right)z}e^{ik_{\theta}R\theta}\mib{u} ,
\end{align}
where $k_{\theta}$ is a wave number in the azimuthal direction,
and $\mib{u}$ is a four-component vector independent of $z$ and $\theta$.
To describe localized modes near $z = 0$, $k_{z}$ must be a complex constant
in each of the two regions of $z > 0$ and $z < 0$
such that $\Im\{k_{z}\} > 0$ for $z > 0$ and $\Im\{k_{z}\} < 0$ for $z < 0$.
The eigenvalue equation is reduced to
\begin{align}
  \left(\mathcal{H}_{\rm BdG}+U_{M}\right)\mib{u} = E\mib{u} ,
\end{align}
where
\begin{align}
  & \mathcal{H}_{\rm BdG} =
           \nonumber \\
  & \left[ \begin{array}{cc}
             \mathcal{A}_{\theta}k_{\theta}\tau_{x}+\mathcal{A}k_{z}\tau_{y}
             -\mu\tau_{0} & \Delta^{\rm eff} \\
             {\Delta^{\rm eff}}^{\dagger}
             & \mathcal{A}_{\theta}k_{\theta}\tau_{x}
               -\mathcal{A}k_{z}\tau_{y} + \mu\tau_{0}
         \end{array}
  \right] .
\end{align}

We consider the case $\mu = 0$, where we can easily obtain
the eigenstates of $\mathcal{H}_{\rm BdG}+U_{M}$ corresponding
to the chiral Majorana modes.
As shown in Appendix~B, the chiral Majorana modes appear
only when $m_{M}^{\rm eff} > \Delta_{0}$ at $\mu = 0$.
The dispersion relation is given as
\begin{align}
     \label{eq:chiral-disp}
   E = -\mathcal{A}_{\theta}k_{\theta} ,
\end{align}
and the corresponding eigenfunction is
\begin{align}
      \label{eq:trial-eigenf_res1}
   \phi(z,\theta)
   = e^{i\left(k_{z}+\frac{\pi}{a}\right)z}e^{ik_{\theta}R\theta}
     \left[ \begin{array}{c}
              1 \\ -1 \\ -1 \\ 1
            \end{array}
     \right]
\end{align}
with
\begin{align}
     \label{eq:def-kz}
   \mathcal{A}k_{z}
   = \left\{ \begin{array}{cc}
               i(m_{M}^{\rm eff}-\Delta_{0}) &  {\rm for} \hspace{3mm} z > 0 \\
               -i\Delta_{0} &  {\rm for} \hspace{3mm} z < 0
             \end{array}
     \right. .
\end{align}

We turn to the case of $\mu \ne 0$.
In this case, we assume from the beginning that the dispersion relation is
$E \propto k_{\theta}$, as in the case of $\mu = 0$,
and restrict our consideration to the limit of $k_{\theta} \to 0$,
where the eigenfunction is obtained in a simple way.
As shown in Appendix~B, we need only consider
the case of $\sqrt{{m_{M}^{\rm eff}}^{2}-\mu^{2}} > \Delta_{0}$.
In the limit of $k_{\theta} \to 0$,
the eigenfunction in the region of $z > 0$ is written as
\begin{align}
      \label{eq:trial-eigenf_res2+}
  \phi(z)
   = c e^{\left(-\kappa + i\frac{\pi}{a}\right)z}
     \left[ \begin{array}{c}
              1 + \frac{m_{M}^{\rm eff}-\sqrt{{m_{M}^{\rm eff}}^{2}-\mu^{2}}}
                       {\mu} \\
             -1 + \frac{m_{M}^{\rm eff}-\sqrt{{m_{M}^{\rm eff}}^{2}-\mu^{2}}}
                       {\mu} \\
             -1 - \frac{m_{M}^{\rm eff}-\sqrt{{m_{M}^{\rm eff}}^{2}-\mu^{2}}}
                       {\mu} \\
              1 - \frac{m_{M}^{\rm eff}-\sqrt{{m_{M}^{\rm eff}}^{2}-\mu^{2}}}
                       {\mu}
            \end{array}
     \right]
\end{align}
with
\begin{align}
  \kappa = \frac{\sqrt{{m_{M}^{\rm eff}}^{2}-\mu^{2}}-\Delta_{0}}
                {\mathcal{A}} .
\end{align}
The eigenfunction in the region of $z < 0$ is written as
\begin{align}
      \label{eq:trial-eigenf_res2-}
  \phi(z)
 & = c \cos\left(\frac{\mu z}{\mathcal{A}}\right)
     e^{\left(\frac{\Delta_{0}}{\mathcal{A}}+ i\frac{\pi}{a}\right)z}
     \left[ \begin{array}{c}
              1 + \frac{m_{M}^{\rm eff}-\sqrt{{m_{M}^{\rm eff}}^{2}-\mu^{2}}}
                       {\mu} \\
             -1 + \frac{m_{M}^{\rm eff}-\sqrt{{m_{M}^{\rm eff}}^{2}-\mu^{2}}}
                       {\mu} \\
             -1 - \frac{m_{M}^{\rm eff}-\sqrt{{m_{M}^{\rm eff}}^{2}-\mu^{2}}}
                       {\mu} \\
              1 - \frac{m_{M}^{\rm eff}-\sqrt{{m_{M}^{\rm eff}}^{2}-\mu^{2}}}
                       {\mu}
            \end{array}
     \right]
         \nonumber \\
 & \hspace{-9mm}
     +c \sin\left(\frac{\mu z}{\mathcal{A}}\right)
     e^{\left(\frac{\Delta_{0}}{\mathcal{A}}+ i\frac{\pi}{a}\right)z}
     \left[ \begin{array}{c}
              -1 + \frac{m_{M}^{\rm eff}-\sqrt{{m_{M}^{\rm eff}}^{2}-\mu^{2}}}
                        {\mu} \\
              -1 - \frac{m_{M}^{\rm eff}-\sqrt{{m_{M}^{\rm eff}}^{2}-\mu^{2}}}
                       {\mu} \\
               1 - \frac{m_{M}^{\rm eff}-\sqrt{{m_{M}^{\rm eff}}^{2}-\mu^{2}}}
                        {\mu} \\
               1 + \frac{m_{M}^{\rm eff}-\sqrt{{m_{M}^{\rm eff}}^{2}-\mu^{2}}}
                        {\mu}
            \end{array}
     \right] .
\end{align}
Equations~(\ref{eq:trial-eigenf_res2+}) and (\ref{eq:trial-eigenf_res2-})
are continuous at $z = 0$.
Taking the limit of $\mu \to 0$, this eigenfunction is reduced to
Eq.~(\ref{eq:trial-eigenf_res1}) with $k_{\theta} = 0$.
Note that an eigenfunction corresponding to the chiral Majorana modes
is obtained only when $\sqrt{{m_{M}^{\rm eff}}^{2}-\mu^{2}} > \Delta_{0}$.
This shows that the chiral Majorana modes appear
when $\mu$ satisfies Eq.~(\ref{eq:condition-mu}).

\section{Appearance of a Majorana Zero Mode}

In accordance with the results in the previous section, we consider that
the chiral Majorana modes with $E(k_{\theta})=-\mathcal{A}_{\theta}k_{\theta}$
appear near the circular edge at $z = 0$.
As shown in Sect.~2, the energy eigenvalues are given as
$E = \mathcal{A}_{\theta}(n+\frac{1}{2})/R$
under the antiperiodic boundary condition.
This shows that there is no Majorana zero mode,
since $k_{\theta} = 0$ is not allowed.

Let us introduce a screw dislocation with the displacement of
$b_{z}$ unit atomic layers piercing the hole.
We show that the antiperiodic boundary condition is transformed into
the periodic boundary condition when $b_{z}$ is an odd integer.
The key is the fact that the Dirac point $(k_{\theta}^{\rm D}, k_{z}^{\rm D})$
of the bogolon states on the inner surface satisfies
$k_{z}^{\rm D} = \frac{\pi}{a}$ according to the weak indices $(001)$.
This leads to the factor $e^{i\frac{\pi}{a}z}$ in $\phi(z,\theta)$
[see, e.g., Eq.~(\ref{eq:trial-eigenf_res1})].
In the absence of the screw dislocation, $\phi(z,\theta)$ must satisfy
the antiperiodic boundary condition
\begin{align}
     \label{eq:a-pdb-phi}
  \phi(z,\theta) = - \phi(z,\theta+2\pi) .
\end{align}
However, once the screw dislocation with $b_{z}$ is introduced,
$\phi(z,\theta)$ must satisfy
\begin{align}
  \phi(z,\theta) = - \phi(z-b_{z}a,\theta+2\pi)
\end{align}
instead of Eq.~(\ref{eq:a-pdb-phi}).
It is convenient to factor out the short wavelength component
$e^{i\frac{\pi}{a}z}$ from $\phi(z,\theta)$ as
\begin{align}
  \phi(z,\theta) = e^{i\frac{\pi}{a}z} \tilde{\phi}(z,\theta) .
\end{align}
The boundary condition for $\tilde{\phi}(z,\theta)$
is expressed as~\cite{imura1,imura2}
\begin{align}
  \tilde{\phi}(z,\theta) = -(-1)^{b_{z}} \tilde{\phi}(z,\theta+2\pi) ,
\end{align}
where the variation of the long wavelength component $\tilde{\phi}(z,\theta)$
on a short length scale is ignored
[i.e., we approximate it as
$\tilde{\phi}(z-b_{z}a,\theta+2\pi) = \tilde{\phi}(z,\theta+2\pi)$].
We find that $\tilde{\phi}(z,\theta)$ obeys the periodic boundary condition
\begin{align}
  \tilde{\phi}(z,\theta) = \tilde{\phi}(z,\theta+2\pi)
\end{align}
when $b_{z}$ is an odd integer.~\cite{comment2}
Under the periodic boundary condition imposed on $\tilde{\phi}(z,\theta)$,
the energy eigenvalues are given as $E = \mathcal{A}_{\theta}n/R$
with $n = 0$, $\pm 1$, $\pm 2$, $\dots$, as shown in Sect.~2. 
This spectrum is equivalent to Eq.~(\ref{eq:spect-with-MZM})
proposed in our scenario.
The eigenvalue $E = 0$ corresponds to a Majorana zero mode.

As discussed in Sect.~2, we can conclude from this spectrum that
the Majorana zero mode appears near the end of a screw dislocation
even in the absence of the hole.

\section{Summary and Discussion}

To explain a dislocation-induced Majorana zero mode
on the surface of an iron-based superconductor,
we have introduced an illustrative model that has a hole parallel
to the $z$-axis, through which a screw dislocation,
designated by the Burgers vector $\mib{b}=(0,0,b_{z})$, is introduced.
Its top surface is assumed to have surface magnetization,
and helical surface states characterized by the weak indices
$(\nu_{x}\nu_{y}\nu_{z}) = (001)$ appear on every surface.
On the basis of this model, we presented a simple scenario to explain
the appearance of a Majorana zero mode near the edge of the hole.
The role of $(\nu_{x}\nu_{y}\nu_{z}) = (001)$ and $\mib{b}=(0,0,b_{z})$ is
explained in terms of a boundary condition for bogolon states.
The bogolon states on the inner surface obey
an antiperiodic boundary condition, which prevents the appearance of
a Majorana zero mode, in the absence of the screw dislocation.
In contrast, the screw dislocation with an odd $b_{z}$ combined with
$(\nu_{x}\nu_{y}\nu_{z}) = (001)$ transforms
the antiperiodic boundary condition into a periodic boundary condition,
allowing the appearance of a Majorana zero mode.
This situation is similar to the appearance of one-dimensional gapless modes
in a topological insulator with a screw dislocation.~\cite{ran,imura1}
The scenario is confirmed by the numerical simulation of a tight-binding model
for iron-based superconductors.
We also present an effective theory that supports the scenario.

\begin{figure}[btp]
\begin{center}
\includegraphics[height=2.0cm]{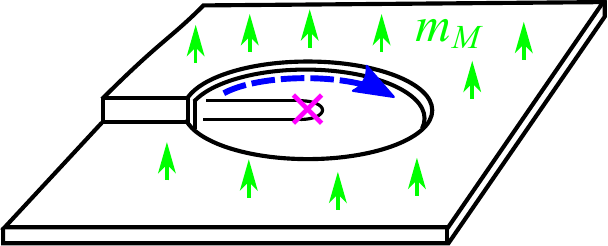}
\end{center}
\caption{
(Color online)
Schematic of the system covered by a ferromagnetic insulating layer
with a hole.
A screw dislocation terminates at the center of the hole,
indicated by the cross.
The dashed arrow represents chiral Majorana modes
localized along the edge of the hole.
}
\end{figure}
In the remainder of this section,
we consider how to create a Majorana zero mode
when there is no magnetization on the top surface of the system.
Constant magnetization can be induced by covering the top surface
with an insulating ferromagnetic layer
so that its magnetization is everywhere perpendicular to the surface.
An experimental difficulty in this case is that we cannot directly access
the Majorana zero mode with a scanning tunneling microscope because
the Majorana zero mode is located under the insulating ferromagnetic layer.
A simple way to avoid this difficulty is to make a small hole
in the insulating ferromagnetic layer such that a screw dislocation terminates
in the center of the hole (see Fig.~4).
In this arrangement, a Majorana zero mode appears along the edge of the hole
if $b_{z}$ is an odd integer.~\cite{comment2}

\section*{Acknowledgment}

The authors thank T. Hanaguri for valuable comments on
several experimental studies on iron-based superconductors.

\appendix

\section{Derivation of $H_{\rm BdG}^{\rm eff}$}

To obtain $H_{\rm BdG}^{\rm eff}$, we derive an effective Hamiltonian
$H_{e}^{\rm eff}$ for the helical surface states of electrons
on the cylindrical inner surface from $H_{e}(\mib{k})$.~\cite{imura4,takane}
Since $(k_{x/y}^{\rm D}, k_{z}^{\rm D})=(0,\frac{\pi}{a})$,
our attention is focused on low-energy states
near $\mib{k}_{0}=(0,0,\frac{\pi}{a})$.
Thus, we approximate $m(\mib{k})$ as the sum of
$m(\mib{k})|_{\mib{k}=\mib{k}_{0}}$ and a small correction to it as
\begin{align}
  m(\mib{k})
  & \approx  m_{0}+ 4m_{\perp}
     \nonumber \\
  & \hspace{-6mm}
            + m_{\parallel}\left((k_{x}a)^{2}+(k_{y}a)^{2}\right)
            - 2m_{\perp}\left( \cos(k_{z}a) + 1 \right) ,
\end{align}
where a long wavelength approximation,
\begin{align}
  \cos(k_{x}a) \approx 1-\frac{(k_{x}a)^{2}}{2} ,
  \hspace{3mm}
  \cos(k_{y}a) \approx 1-\frac{(k_{y}a)^{2}}{2} ,
\end{align}
is used.
For later convenience, we decompose $m(\mib{k})$ as
\begin{align}
  m(\mib{k})
  = m_{xy}(k_{x},k_{y}) - 2m_{\perp}\left( \cos(k_{z}a) + 1 \right)
\end{align}
with
\begin{align}
  m_{xy}(k_{x},k_{y})
  = m_{0}+ 4m_{\perp} + m_{\parallel}a^{2}\left(k_{x}^{2}+k_{y}^{2}\right) .
\end{align}
We also approximate $p_{\pm}(k_{x},k_{y})$ as
\begin{align}
  p_{\pm}(k_{x},k_{y}) \approx a \left(k_{x} \pm i k_{y}\right) .
\end{align}
Then, we rewrite $m_{xy}(k_{x},k_{y})$ and $p_{\pm}(k_{x},k_{y})$
in terms of $\theta$ and the radial coordinate $r$ ($r \ge R$) as
\begin{align}
  m_{xy}
  & = M_{0}
      + M_{2}\left(-\frac{1}{r^{2}}\frac{\partial^{2}}{\partial \theta^{2}}
                   -\frac{1}{r}\frac{\partial}{\partial r}
                   -\frac{\partial^{2}}{\partial r^{2}}
             \right) ,
           \\
  p_{\pm}
  & = a e^{\pm i\theta}\left( \pm\frac{1}{r}\frac{\partial}{\partial \theta}
                              -i \frac{\partial}{\partial r} \right) ,
\end{align}
where $M_{0} = m_{0}+4m_{\perp}$ and $M_{2}=m_{\parallel}a^{2}$.
We express the resulting real-space representation of $H_{e}(\mib{k})$
as $H_{e}(r,\theta)$.

To find eigenstates of $H_{e}(r,\theta)$, we decompose $H_{e}(r,\theta)$
as $H_{e}(r,\theta) = H_{\perp}+H_{\parallel}$ with
\begin{align}
  H_{\perp}
  & = \left[ \begin{array}{cc}
               \left(M_{0}-M_{2}\hat{\xi}\right)\tau_{z} 
               & -i\mathcal{A}e^{-i\theta}\frac{\partial}{\partial r}\tau_{x}\\
               -i\mathcal{A}e^{i\theta}\frac{\partial}{\partial r}\tau_{x}
               & \left(M_{0}-M_{2}\hat{\xi}\right)\tau_{z}\\
             \end{array}
      \right] ,
        \\
  H_{\parallel}
  & = \left[ \begin{array}{cc}
               \hat{\Gamma}\tau_{z}+A\sin(k_{z}a)\tau_{x}
               & \mathcal{A}\hat{k}_{-}\tau_{x}\\
               \mathcal{A}\hat{k}_{+}\tau_{x}
               & \hat{\Gamma}\tau_{z}-A\sin(k_{z}a)\tau_{x}
             \end{array}
      \right] ,
\end{align}
where $\mathcal{A}=Aa$, and
\begin{align}
  \hat{\xi}
  & = \frac{1}{r}\frac{\partial}{\partial r}
      + \frac{\partial^{2}}{\partial r^{2}} ,
          \\
  \hat{\Gamma}
  & = -2m_{\perp}\left(\cos(k_{z}a)+1\right)
      -M_{2}\frac{1}{R^{2}}\frac{\partial^{2}}{\partial \theta^{2}} ,
          \\
  \hat{k}_{\pm}
  & = \pm e^{\pm i \theta}\frac{1}{R}\frac{\partial}{\partial \theta} .
\end{align}
Note that $H_{\perp}$ describes the penetration of surface electron states
into the bulk ($r > R$), whereas $H_{\parallel}$ describes the dispersion
relation of the surface electron states.

Let us solve the eigenvalue equation for $H_{\perp}$ using a trial function
given by
\begin{align}
  |\psi(r,\theta)\rangle = e^{-\kappa(r-R)}|\varphi(\theta)\rangle ,
\end{align}
where $\kappa$ is a positive constant characterizing the penetration of
a surface electron state.
For this trial function, the eigenvalue equation is simplified to
\begin{align}
    \label{eq:eigv-eq_perp}
 &   \left[ \begin{array}{cc}
              \left(M_{0}-M_{2}\xi(\kappa)\right)\tau_{z} 
              & i\mathcal{A}e^{-i\theta}\kappa\tau_{x}\\
              i\mathcal{A}e^{i\theta}\kappa\tau_{x}
              & \left(M_{0}-M_{2}\xi(\kappa)\right)\tau_{z}
            \end{array}
     \right]
     |\varphi(\theta)\rangle
         \nonumber \\
 & = E_{\perp}|\varphi(\theta)\rangle
\end{align}
with
\begin{align}
  \xi(\kappa) = \kappa^{2} - \frac{\kappa}{R} .
\end{align}
The eigenvalue $E_{\perp}$ satisfies
\begin{align}
    \label{eq:det-H_perp}
  & \{E_{\perp}^{2}-\left(M_{0}-M_{2}\xi(\kappa)\right)^{2}\}^{2}
    + \left(\mathcal{A}^{2}\kappa^{2}\right)^{2}
       \nonumber \\
  & \hspace{4mm}
    + 2\mathcal{A}^{2}\kappa^{2}
      \{E_{\perp}^{2}-\left(M_{0}-M_{2}\xi(\kappa)\right)^{2}\}
     = 0 .
\end{align}
We construct an eigenfunction of $H_{\perp}$ by superposing two solutions
with $\kappa_{+}$ and $\kappa_{-}$ as
\begin{align}
     \label{eq:eigenf-penet}
 |\psi(r,\theta)\rangle
   = a_{+}e^{-\kappa_{+}(r-R)}|\varphi_{+}(\theta)\rangle
   + a_{-}e^{-\kappa_{-}(r-R)}|\varphi_{-}(\theta)\rangle ,
\end{align}
where $a_{+}$ and $a_{-}$ are constants.
Note that $|\Psi(r,\theta)\rangle$ must satisfy the open boundary condition
on the surface, i.e.,
\begin{align}
 |\Psi(R,\theta)\rangle = \mib{0} .
\end{align}
This is satisfied only when $\kappa_{+} \neq \kappa_{-}$ and
\begin{align}
  |\varphi_{+}(\theta)\rangle = |\varphi_{-}(\theta)\rangle .
\end{align}
We can show that this holds only when the two equations,
\begin{align}
    \label{eq:two-equations_1}
   \frac{M_{0}-M_{2}\xi_{+}-E_{\perp}}{i\mathcal{A}\kappa_{+}}
 &
   = \frac{M_{0}-M_{2}\xi_{-}-E_{\perp}}{i\mathcal{A}\kappa_{-}},
        \\
    \label{eq:two-equations_2}
   \frac{-M_{0}+M_{2}\xi_{+}-E_{\perp}}{i\mathcal{A}\kappa_{+}}
 &
   = \frac{-M_{0}+M_{2}\xi_{-}-E_{\perp}}{i\mathcal{A}\kappa_{-}}
\end{align}
with $\xi_{\pm} \equiv \xi(\kappa_{\pm})$,
are simultaneously satisfied.~\cite{takane}
These equations are easily obtained
if Eq.~(\ref{eq:eigv-eq_perp}) is rewritten as
\begin{align}
    \label{eq:eigv-eq_red}
 &   \left[ \begin{array}{cc}
              \frac{\left(M_{0}-M_{2}\xi_{\pm}\right)\tau_{z}
                    -E_{\perp}\tau_{0}}
                   {i\mathcal{A}\kappa_{\pm}}
              & e^{-i\theta}\tau_{x}\\
              e^{i\theta}\tau_{x}
              & \frac{\left(M_{0}-M_{2}\xi_{\pm}\right)\tau_{z}
                      -E_{\perp}\tau_{0}}
                     {i\mathcal{A}\kappa_{\pm}}
            \end{array}
     \right]
     |\varphi_{\pm}(\theta)\rangle
         \nonumber \\
 & = \mib{0} .
\end{align}
Equations~(\ref{eq:two-equations_1}) and (\ref{eq:two-equations_2}) result in
$E_{\perp} = 0$.
Combining this with Eq.~(\ref{eq:det-H_perp}), we find
$M_{0}-M_{2}\xi(\kappa) = \chi \mathcal{A}\kappa$, where $\chi = +$ or $-$.
In our case of $M_{0} < 0$, $M_{2} > 0$, and $\mathcal{A} > 0$,
the proper solutions are obtained in the case of $\chi = -$ as
\begin{align}
  \kappa_{\pm}
  = \frac{\frac{M_{2}}{R}+\mathcal{A}
          \pm\sqrt{\left(\frac{M_{2}}{R}+\mathcal{A}\right)^{2}+4M_{0}M_{2}}}
         {2M_{2}}
\end{align}
with the two eigenvectors
$|\varphi_{e1}(\theta)\rangle$ and $|\varphi_{e2}(\theta)\rangle$
given as Eqs.~(\ref{eq:def-basis-v1}) and (\ref{eq:def-basis-v2}).
These two eigenvectors,
each of which satisfies Eq.~(\ref{eq:eigv-eq_red}) with $E_{\perp} = 0$,
serve as the basis vectors in the electron space.

Using $|\varphi_{e1}(\theta)\rangle$ and $|\varphi_{e2}(\theta)\rangle$,
we derive an effective Hamiltonian $H_{e}^{\rm eff}$
for the helical surface states of electrons from $H_{\parallel}$.
In the lowest-order approximation with respect to $H_{\parallel}$,
$H_{e}^{\rm eff}$ is given as
\begin{align}
     \label{eq:H_e-def}
  H_{e}^{\rm eff}
  = \left[ \begin{array}{cc}
             D_{11} & D_{12} \\
             D_{21} & D_{22}
           \end{array}
    \right] ,
\end{align}
where
\begin{align}
  D_{ij} = \langle\varphi_{e i}(\theta)|H_{\parallel}
           |\varphi_{e j}(\theta)\rangle
\end{align}
with $i, j = 1, 2$.
We easily find that
\begin{align}
     \label{eq:D11}
   D_{11} & = D_{22} = 0 ,
      \\
   D_{12} & = \left(Aa+\frac{m_{\parallel}a}{R}\right)\hat{k}_{\theta}
              +iA\sin(k_{z}a) ,
      \\
     \label{eq:D21}
   D_{21} & = \left(Aa+\frac{m_{\parallel}a}{R}\right)\hat{k}_{\theta}
              -iA\sin(k_{z}a) ,
\end{align}
where $\hat{k}_{\theta}$ is defined in Eq.~(\ref{eq:def-k_theta-z}).
Since $H_{e}^{\rm eff}$ describes low-energy states near
$\mib{k}_{0}=(0,0,\frac{\pi}{a})$,
we use the long wavelength approximation
$\sin(k_{z}a) \approx -a \left(k_{z} - \frac{\pi}{a}\right)$
and then replace $k_{z}$ with $\hat{k}_{z}$
defined in Eq.~(\ref{eq:def-k_theta-z}).
This results in
\begin{align}
     \label{eq:k_z-shift}
  \sin(k_{z}a) \approx -a \left(\hat{k}_{z} - \frac{\pi}{a}\right) .
\end{align}
Substituting Eqs.~(\ref{eq:D11})--(\ref{eq:D21}) with Eq.~(\ref{eq:k_z-shift})
into Eq.~(\ref{eq:H_e-def}),
we arrive at the expression of $H_{e}^{\rm eff}$
given as Eq.~(\ref{eq:H_e_eff}).

Similarly, we can derive $H_{h}^{\rm eff}$ given as Eq.~(\ref{eq:H_h_eff})
from $H_{h}(\mib{k})$.
We do not describe this derivation.
The basis vectors
$|\varphi_{h 1}(\theta)\rangle$ and $|\varphi_{h 2}(\theta)\rangle$
in the hole space
are given as Eqs.~(\ref{eq:def-basis-v3}) and (\ref{eq:def-basis-v4}).

Finally, we derive $\Delta^{\rm eff}$ given as Eq.~(\ref{eq:Delta_eff})
from $\Delta$ given as Eq.~(\ref{eq:Delta}).
The Fourier transform of $\Delta$ does not depend on $\mib{k}$
and is simply given by
\begin{align}
 \Delta(\mib{k})
 = \left[ \begin{array}{cc}
            0 & \Delta_{0}\tau_{0} \\
            -\Delta_{0}\tau_{0} & 0
          \end{array}
   \right] .
\end{align}
The effective pair potential $\Delta^{\rm eff}$ is expressed as
\begin{align}
     \label{eq:Delta_e-def}
 \Delta^{\rm eff}
 = \left[ \begin{array}{cc}
            \Delta_{11} & \Delta_{12} \\
            \Delta_{21} & \Delta_{22}
          \end{array}
   \right] ,
\end{align}
where
\begin{align}
  \Delta_{ij} = \langle\varphi_{e i}(\theta)|\Delta(\mib{k})
                |\varphi_{h j}(\theta)\rangle
\end{align}
with $i, j  = 1, 2$.
We easily find
\begin{align}
   \Delta_{11} = \Delta_{22} = 0 ,
   \hspace{5mm}
   \Delta_{21} = - \Delta_{12} = \Delta_{0} .
\end{align}
Substituting these equations into Eq.~(\ref{eq:Delta_e-def}),
we arrive at the expression of
$\Delta^{\rm eff}$ given as Eq.~(\ref{eq:Delta_eff}).

\section{Derivation of Chiral Majorana Modes}

Let us consider the case $\mu = 0$, where
${\rm det}\{\mathcal{H}_{\rm BdG}+U_{M}-E\} = 0$ gives
\begin{align}
  \mathcal{A}k_{z} =\pm i \sqrt{\left(m_{M}(z)\pm\Delta_{0}\right)^{2}
                                -E^{2}+(\mathcal{A}_{\theta}k_{\theta})^{2}}
\end{align}
if $\left(m_{M}(z)\pm\Delta_{0}\right)^{2}
> E^{2}-(\mathcal{A}_{\theta}k_{\theta})^{2}$.
As noted in Sect.~4, $k_{z}$ must satisfy $\Im\{k_{z}\} > 0$ for $z > 0$
and $\Im\{k_{z}\} < 0$ for $z < 0$.
From this observation, we can set $k_{z}$ as
\begin{align}
  \mathcal{A}k_{z} = \left\{ \begin{array}{cc}
                             i\sqrt{\left(m_{M}^{\rm eff}\pm\Delta_{0}
                                    \right)^{2}
                                    -E^{2}+(\mathcal{A}_{\theta}k_{\theta})^{2}}                             &  {\rm for} \hspace{3mm} z > 0 \\
                            -i\sqrt{\Delta_{0}^{2}
                                    -E^{2}+(\mathcal{A}_{\theta}k_{\theta})^{2}}                             &  {\rm for} \hspace{3mm} z < 0
                           \end{array}
                   \right..
\end{align}
The eigenvector $\mib{u}$ in Eq.~(\ref{eq:trial-eigenf})
must be independent of $z$.
Examining this requirement,
we finally find that $\mib{u}$ becomes independent of $z$ as
$\mib{u} = {}^{t}\!\left[ +1 -1 -1 +1 \right]$ when
\begin{align}
   E = -\mathcal{A}_{\theta}k_{\theta}
\end{align}
with
\begin{align}
   \mathcal{A}k_{z}
   = \left\{ \begin{array}{cc}
               i(m_{M}^{\rm eff}-\Delta_{0}) &  {\rm for} \hspace{3mm} z > 0 \\
               -i\Delta_{0} &  {\rm for} \hspace{3mm} z < 0
             \end{array}
     \right. ,
\end{align}
where $m_{M}^{\rm eff} > \Delta_{0}$ is assumed.
The resulting eigenfunction is given as Eq.~(\ref{eq:trial-eigenf_res1}).
If $m_{M}^{\rm eff} < \Delta_{0}$, the eigenstates corresponding to
the chiral Majorana modes do not appear.

We turn to the case of $\mu \ne 0$.
In this case, we restrict our consideration to the limit of $k_{\theta} \to 0$,
where the eigenfunction is obtained in a simple way.
In the limit of $k_{\theta} \to 0$,
${\rm det}\{\mathcal{H}_{\rm BdG}+U_{M}-E\} = 0$ yields
\begin{align}
  \mathcal{A}k_{z} = i \zeta \left(\sqrt{{m_{M}^{\rm eff}}^{2}
                                   -\mu^{2}}-\Delta_{0}\right)
\end{align}
for $z > 0$, where $\zeta = +$ or $-$.
Since $\Im\{k_{z}\} > 0$ for $z > 0$, we must choose $\zeta = +$
if $\sqrt{{m_{M}^{\rm eff}}^{2}-\mu^{2}} > \Delta_{0}$
and $\zeta = -$ if $\sqrt{{m_{M}^{\rm eff}}^{2}-\mu^{2}} < \Delta_{0}$.
At $\mu = 0$, we find that the chiral Majorana modes appear
only when $m_{M}^{\rm eff} > \Delta_{0}$.
With this in mind, we are allowed to consider
only the case of $\sqrt{{m_{M}^{\rm eff}}^{2}-\mu^{2}} > \Delta_{0}$.
The eigenvector corresponding to the case of $\zeta = +$ is determined as
\begin{align}
     \label{eq:def-u_z>0}
 \mib{u}
 = \left[ \begin{array}{c}
            1 + \frac{m_{M}^{\rm eff}-\sqrt{{m_{M}^{\rm eff}}^{2}-\mu^{2}}}
                     {\mu} \\
           -1 + \frac{m_{M}^{\rm eff}-\sqrt{{m_{M}^{\rm eff}}^{2}-\mu^{2}}}
                     {\mu} \\
           -1 - \frac{m_{M}^{\rm eff}-\sqrt{{m_{M}^{\rm eff}}^{2}-\mu^{2}}}
                     {\mu} \\
            1 - \frac{m_{M}^{\rm eff}-\sqrt{{m_{M}^{\rm eff}}^{2}-\mu^{2}}}
                     {\mu}
          \end{array}
   \right] ,
\end{align}
where a normalization constant is ignored.
For $z < 0$, we find that $k_{z}$ is determined as
\begin{align}
     \label{eq:def-k_z-pm}
  \mathcal{A}k_{z}^{\pm} = \pm \mu -i \Delta_{0}
\end{align}
and the eigenvector corresponding to it is
\begin{align}
 \mib{u}_{\pm}
 = \left[ \begin{array}{c}
            1 \pm i \\
           -1 \pm i \\
           -1 \mp i \\
            1 \mp i
            \end{array}
   \right] ,
\end{align}
where a normalization constant is also ignored.

We examine whether the solutions in the two regions of $z > 0$ and $z < 0$
can be connected at $z = 0$.
In the region of $z > 0$, the solution with $k_{\theta} = 0$ is given as
\begin{align}
  \phi(z)
   = c \, e^{\left(-\kappa + i\frac{\pi}{a}\right)z}\mib{u} ,
\end{align}
where $c$ is a normalization constant,
$\mib{u}$ is given as Eq.~(\ref{eq:def-u_z>0}), and
\begin{align}
  \mathcal{A}\kappa = \sqrt{{m_{M}^{\rm eff}}^{2}-\mu^{2}}-\Delta_{0} .
\end{align}
This is identical to Eq.~(\ref{eq:trial-eigenf_res2+}).
In the region of $z < 0$, the solution with $k_{\theta} = 0$ is expressed
in the form of
\begin{align}
  \phi(z)
   = a \, e^{i\left(k_{z}^{+}+ \frac{\pi}{a}\right)z}
     \left[ \begin{array}{c}
              1 + i \\
             -1 + i \\
             -1 - i \\
              1 - i
            \end{array}
     \right]
    +b \, e^{i\left(k_{z}^{-}+ \frac{\pi}{a}\right)z}
     \left[ \begin{array}{c}
              1 - i \\
             -1 - i \\
             -1 + i \\
              1 + i
            \end{array}
     \right]
\end{align}
with $k_{z}^{\pm}$ given in Eq.~(\ref{eq:def-k_z-pm}).
These two solutions are connected at $z = 0$ if we set
\begin{align}
  a & = c\left(\frac{1}{2}
               -i\frac{m_{M}^{\rm eff}-\sqrt{{m_{M}^{\rm eff}}^{2}-\mu^{2}}}
                      {2\mu}\right) ,
          \\
  b & = c\left(\frac{1}{2}
               +i\frac{m_{M}^{\rm eff}-\sqrt{{m_{M}^{\rm eff}}^{2}-\mu^{2}}}
                      {2\mu}\right) .
\end{align}
The resulting eigenfunction is given as Eq.~(\ref{eq:trial-eigenf_res2-}).


\begin{thebibliography}{99}
\bibitem{comment0} In a general case of $(\nu_{x}\nu_{y}\nu_{z}) \neq (000)$,
a Majorana zero mode appears in the presence of a screw dislocation designated
by $\mib{b}=(b_{x},b_{y},b_{z})$ when $\mib{b} \cdot \mib{M}_{\rm w} = 1$
$({\rm mod}\, 2)$, where $\mib{M}_{\rm w}=(\nu_{x},\nu_{y},\nu_{z})$.
This condition is equivalent to the one discussed in Ref.~~\citen{ran}.
In the case of $\mib{b} \parallel \hat{z}$,
$\nu_{x}$ and $\nu_{y}$ are irrelevant.
\bibitem{comment1} The twofold degeneracy is caused by
the two equivalent cuts.
\bibitem{comment2} When a wavefunction on the top surface of
the original setup obeys an antiperiodic boundary condition
along a closed path encircling the hole, the antiperiodic boundary condition
is transformed into a periodic one if a screw dislocation with an odd $b_{z}$
is inserted  through the hole
because a phase shift $\pi$ is induced across the side surface of
height $b_{z}$, on which $k_{z}^{\rm D} = \frac{\pi}{a}$.
\end{thebibliography}
\end{document}